\documentclass{jfm}
\usepackage{graphicx}
\usepackage{epstopdf, epsfig}
\usepackage{soul}
\usepackage[normalem]{ulem}

\usepackage{color}

\shorttitle{Data-driven Control Method for Impinging Jets}
\shortauthor{Spencer L. Stahl and Datta V. Gaitonde}

\title{A data-driven approach to guide supersonic impinging jet control}

\author{Spencer L. Stahl\aff{1}
  \corresp{\email{stahl.174@osu.edu}} \and
  Datta V. Gaitonde\aff{1}}

\affiliation{\aff{1}Department of Mechanical and Aerospace Engineering, The Ohio State University, OH 43210, USA}

\begin{document}

\maketitle

\begin{abstract}

A data-driven framework using snapshots of an uncontrolled flow is proposed to identify, and subsequently demonstrate, effective control strategies for different objectives in supersonic impinging jets.
The approach, based on a dynamic mode decomposition reduced order model (DMD-ROM), determines forcing receptivity in an economical manner by projecting flow and actuator-specific forcing snapshots onto a reduced subspace and then evolving the results forward in time.
Since it effectively determines a linear response around the unsteady flow in the time-domain, the method differs materially from typical techniques that use steady basic states, such as stability or input-output approaches that employ linearized Navier-Stokes operators in the frequency-domain.
The method naturally accounts for factors inherent to the snapshot basis, including configuration complexity and flow parameters such as Reynolds number.
Furthermore, gain metrics calculated in the reduced subspace facilitate rapid assessments of flow sensitivities to a wide range of forcing parameters, from which optimal actuator inputs may be selected and results confirmed in scale-resolved simulations or experiments.
The DMD-ROM approach is demonstrated from two different perspectives.
The first concerns asymptotic feedback resonance, where the effects of harmonic pressure forcing are estimated and verified with nonlinear simulations using a blowing-suction actuator.
The second examines time-local behavior within critical feedback events,
where the phase of actuation becomes important.
For this, a conditional space-time mode is used to identify the optimal forcing phase that minimizes convective instability initiation within the resonance cycle.

\end{abstract}

\section{Introduction}
Active flow control has the potential to alleviate many aerodynamic problems; some objectives include mitigation of thermomechanical loads, separation and unsteadiness, or flow-acoustic interactions due to supersonic jets impinging on a ground plane, of interest in this work.
Optimal control inputs are predicated on identifying responsiveness to external forcing (receptivity to perturbations) that depend on the underlying flow mechanisms, which are often examined with the linearized frequency-domain Navier-Stokes equations applied to steady basic states.
Classical linear stability extracts such information from the laminar state \citep{Theofilis2011,Juniper2014}, or, the time-averaged turbulent state, whose modes are related to larger turbulent coherent structures \citep{crighton1976stability}.
More recently, other manners of analyzing the mean flow have been employed with overlapping objectives to extract singular or eigenvalue spectra with corresponding modes \citep{Schmid2007, Luchini2014_adjoint,Ranjan2020_mfp}.
From a control standpoint, optimal forcing-response relationships may be derived from the mean flow through resolvent or input-output modes that maximize linear energy growth.
Example applications 
of control strategies derived from perturbed steady basic states may be found in \citet{Herrmann2021,herrmann2023resolvent}. 



When unsteady data is available in the form of snapshots, additional information may be derived to discern the key mechanisms as well as to propose control inptus.
Modal decomposition techniques \citep{Taira2020_modal_b} have become increasingly popular to represent the dynamics (model order reduction),  and their
potential implications for control have also been elaborated on \citep{Rowley2017_control,Brunton2015}.  
In the present work, we propose a data-driven approach (\S\ref{secn:method}) to derive a forcing-response relationship using perturbations in the time-varying flow with 
snapshot-based modal decomposition serving as a linear surrogate for the dynamics.
Dynamic Mode Decomposition (DMD) provides such a representation \citep{Schmid2010} and is thus used to generate an efficient reduced order model (DMD-ROM) for the nonlinear behavior contained in the snapshots. 
The method also provides a reduced basis for projection of the input control forcing and the evolution in time of the forced system.
The procedure may be viewed as a much simplified data-driven analog of the synchronized Large Eddy Simulation (LES) method of \citet{adler2018dynamic} to determine time-local perturbation growth in the turbulent flow.

Several advantages may be delineated.
Since it is data-driven, the DMD-ROM contains the main dynamics encapsulated in the snapshots, including configuration complexity and flow parameters such as Mach and Reynolds numbers, large values of which can often impede more difficult linearized Navier-Stokes driven approaches because of spurious modes.
In addition, the time-domain nature of the method eases specification of realizable actuator-specific inputs, including the nature of perturbations, localization in space, and phase variations in time, this last advantage also enables control assessments of time-local events.
The approach may also be applied only with select snapshot variables; this is advantageous, for example, when  data is obtained from experiments.
More crucially, even when all data is available, say, from uncontrolled scale-resolved simulations, only a subset of variables may be selected to construct the DMD-ROM.
In the impinging jet problem of interest, the hydrodynamic-acoustic interaction is well encapsulated in the scale-resolved pressure field, and is therefore the only variable used.


For computational economy -- an important consideration for time-domain methods -- we use a reduced subspace, constructed from projection of flow and forcing snapshots onto a Proper Orthogonal Decomposition (POD) basis.
This
significantly speeds up parametric studies of actuator locations and forcing properties.
Evaluation metrics such as energy gain for asymptotic or short-time events are also computed in the reduced space; projection back to the full physical space may be performed only when needed.

The capabilities of the DMD-ROM approach are highlighted in the context of a complex supersonic impinging jet containing interacting shocks and nonlinear instabilities.
The resonant dynamics have been extensively reviewed by \citet{Edgington-Mitchell2019_review}.
Briefly, Kelvin-Helmholtz structures in the shear-layer emanating from the nozzle exit impinge on the ground plane to generate acoustic waves that travel back to the nozzle exit to complete the loop. 
Two applications of the DMD-ROM are considered to address asymptotic and short-time phenomena, respectively.
The first examines the tonal behavior of the impinging jet.
For this, we employ harmonic pressure forcing at the nozzle receptivity region to interfere with the aeroacoustic feedback resonance and modify the most amplified tones. 
The forced linear response is confirmed with \textit{post facto} scale-resolved simulations featuring realizable blowing-suction actuators with the same spectral forcing properties.

The second aspect concerns transient or time-local control to address the problem of energy growth intrinsic to conditionally selected events occurring over finite time. 
This example targets the vital acoustic feedback receptivity events in the resonance cycle which initiate the convective shear-layer instability.
Perturbation-based insights into such transient structures may be found in \citet{Karami2018}.
Here we use conditional space-time proper orthogonal decomposition (CST-POD) method to isolate an ensemble of shear-layer events directly from the unsteady flow 
\citep{Schmidt2019,stahl_2023_CPOD_jcp} for perturbation-based receptivity analysis, with the objective of determining the optimal forcing phase to mitigate the convective instability that perpetuates the resonance cycle.

\section{Time-Domain DMD-ROM Forcing Methodology}\label{secn:method}





\begin{figure}
\centering
\includegraphics[width=0.9\textwidth]{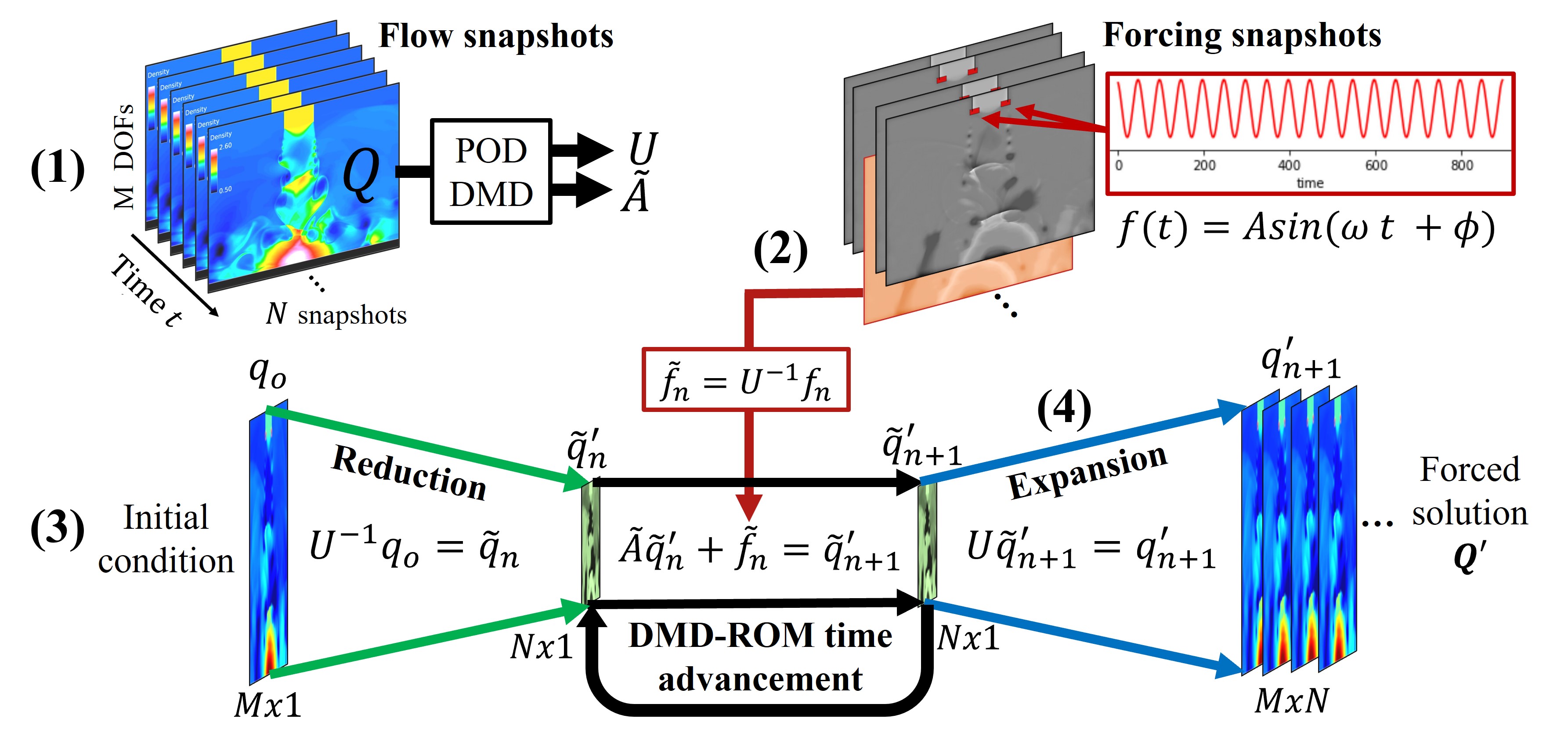}
\caption{Elements of the DMD control method. Step (1) calculation of the DMD operator from the unforced data. Step (2) projection of forcing snapshots onto POD modes. Step (3) overall schematic including initial condition projection and time iteration scheme in the reduced space and (4) projection back to physical space.}
\label{fig:DMDC_method}
\vspace*{-0.1in}
\end{figure}

The proposed method, summarized in the schematic of Fig.~\ref{fig:DMDC_method}, is implemented in four steps: (1) construction of the DMD-ROM from snapshots of the unforced flow, (2) creation and projection on the reduced space of the forcing snapshot time-sequence, (3) 
time advancement of the linearized forcing response and gain evaluation in the reduced space and (4) expansion of the forced response back to the full-space when desired.

The first step (1), uses the unforced sequence of $N$ flow snapshots, $\textbf{Q}=[q_1,...,q_N]$, to model the linear operator $\textbf{A}$ that marches the system from one snapshot to the next, $q_{n+1}=\textbf{A}q_n$. Since $\textbf{Q}$ and $\textbf{A}$ are full sized, $M \times N$ and $M \times M$ matrices, where $M$ is the number of spatial degrees of freedom, 
the classical DMD approach \citep{Schmid2010} is employed to obtain the reduced $N \times N$ operator, $\tilde{\textbf{A}}$, by projection onto the POD modes, $\textbf{U}$, obtained from the singular value decomposition of $\textbf{Q}$.

The second step (2), constructs the $M\times N$-sized forcing snapshot matrix, $\textbf{F}=[f_1,...,f_N]$, based on the chosen actuator location and perturbation properties.
For the jet examples, the most natural location for actuators is the region around the nozzle exit.
Although the forcing function may be an arbitrary function of time, here we consider harmonic forcing $f(t)=A\sin(\omega t + \phi)$, with $A$, $\omega$ and $\phi$ being the amplitude, angular frequency and phase shift in time, respectively. 
The reduced flow and forcing snapshots are projected onto the POD modes of the flow, $\tilde{q}_n=\textbf{U}^{-1}q_n$ and $\tilde{f_n}=\textbf{U}^{-1}f_n$, respectively. 

The third step (3), shown in the overall scheme of Fig.~\ref{fig:DMDC_method}, evolves the reduced forced system in time,
\begin{equation}\label{eqn:dmd_Atild_control}
    \tilde{q}'_{n+1}=\tilde{\textbf{A}} \tilde{q}'_{n} +  \tilde{f}_{n}, \hspace*{0.3in} n=0,1,\ldots,N
\end{equation}
where $\tilde{q}'_{n+1}$ represents the linear response of the system and $n=0$ represents the initial condition. 
Since interpreting the forced results directly from the reduced coordinates can be ambiguous, in the fourth step (4), the full space equivalent of $q'_{n+1}$ may be reassembled when desired by projection back to the full-space domain to obtain the forced solution ($\textbf{Q}'=\textbf{U}\tilde{\textbf{Q}}'$).
For the present feed-forward study, projection back to the full-space is only necessary after the time-iteration stage is complete. 
Full-space expansion within each iteration may become necessary in feedback control to inform changes in the forcing.



The procedure to obtain gain in the reduced space exploits features that are consistent between it and the full-sized system.
Specifically, we use the $L2$-norm ratio of the forced and unforced solutions  $\sigma=||\tilde{\textbf{Q}}'||_2/||\tilde{\textbf{Q}}||_2$ to obtain  a relative sense of gain.
This provides an adequate assessment for the asymptotic forcing analysis (\S\ref{sec:examp1}).
For the transient problem (\S\ref{secn:examp2}), we use the ratio of the $L2$-norm of individual snapshots to the initial condition snapshot: $\sigma(t)=||\tilde{q}(t)||_2/||\tilde{q}_o||_2$.
Since this transient norm applies to both unforced and forced systems, the value may then be assessed to understand the relative effects of the forcing over time.
Operations in the reduced space assure computational economy and enable testing of a wide range of forcing parameters.


Finally, we note that the approach differs from the DMD Control (DMDC) method of \citet{Rosenfeld2021}, which uses forced data for system identification to distinguish the dynamic and forcing operators.
The current DMD-ROM approach could potentially be enriched with such data when combined with a learned control operator; however the focus here is on obtaining the linear response of the unforced dataset to different forcing parameters, optimal results from which may be refined with limited computationally intense nonlinear simulations or experimental testing.

\section{Application to Supersonic Impinging Jets}\label{secn:control}
\begin{figure}
\centering
\includegraphics[width=0.9\textwidth]{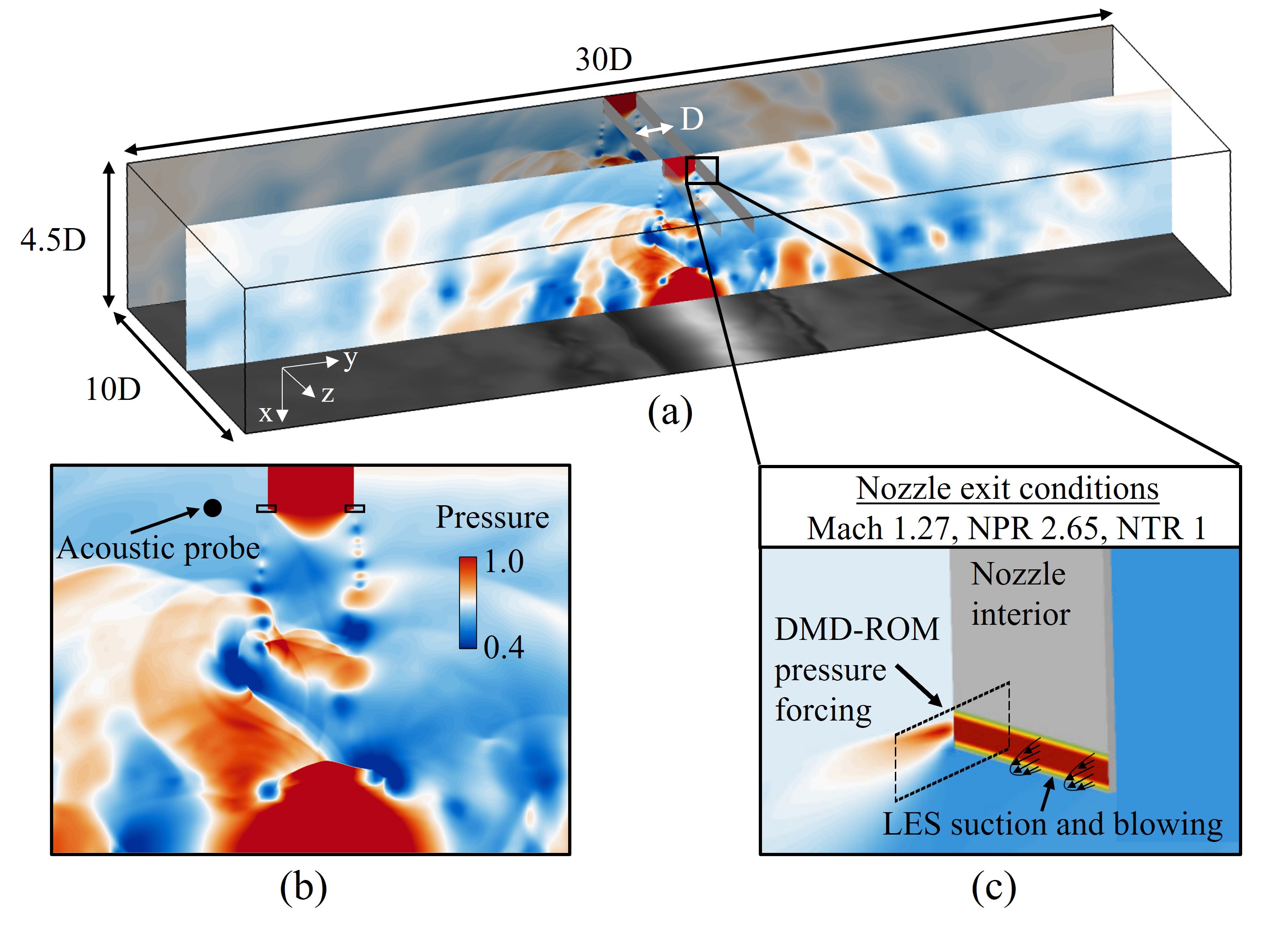}
\vspace*{-0.1in}
\caption{(a) Planar impinging jet domain. (b) Instantaneous uncontrolled pressure snapshot and DMD-ROM initial condition with acoustic probe location. (c) nozzle forcing location (rectangle) for the DMD-ROM (pressure) and LES (blowing-suction).}
\label{fig:SIJ_intro}
\vspace*{-0.1in}
\end{figure}

We consider the aeroacoustic resonance of a planar (spanwise homogeneous), Mach~1.27, underexpanded jet (nozzle-width, $D=0.0254m$) impinging on a plate surface located $4D$ away, as shown in Fig.~\ref{fig:SIJ_intro}(a) which also displays domain dimensions.
The uncontrolled (baseline) data, containing multiple resonant tones, is acquired from an LES at a Reynolds number of $Re=5.8\times10^{5}$ and comprises $3{,}855$  pressure snapshots from the mid-plane of the domain sampled at a nondimensional frequency $St=fD/a=10$, where $f$ is the frequency (Hz) and $a=343 m/s$ is the ambient speed of sound.
The methodology for the LES, including numerical scheme and mesh, follow those established in \citet{stahl_caes2022}.

As noted earlier, the pressure variable is  suitable for the dynamics of interest since it encompasses the hydrodynamic and acoustic phenomena that dictate the feedback loop.
A DMD-ROM based on mean-subtracted pressure fluctuations aids in stability by ensuring all DMD eigenvalues are neutrally stable \citep{Towne2018_relation}.
A representative baseline pressure snapshot, shown in Fig.~\ref{fig:SIJ_intro}(b), 
illustrates the natural asymmetric (at this time instant, predominantly blue on the right, red on the left outside the stagnation region) side-to-side flapping of the jet and feedback acoustics.
The probe location in Fig.~\ref{fig:SIJ_intro}(b) monitors the acoustic feedback tones used in the rest of the paper. 
The forcing is applied on rectangular regions of size $0.2D \times 0.07D$ centered at the nozzle lip exit, displayed in the inset of Fig.~\ref{fig:SIJ_intro}(c), which shows both pressure forcing for the DMD-ROM and blowing-suction actuation for the LES.

\subsection{Example 1: Asymptotic forcing analysis} \label{sec:examp1}
The first example illustrates the asymptotic tonal response of the forcing model.
\begin{figure}
\centering
\includegraphics[width=0.9\textwidth]{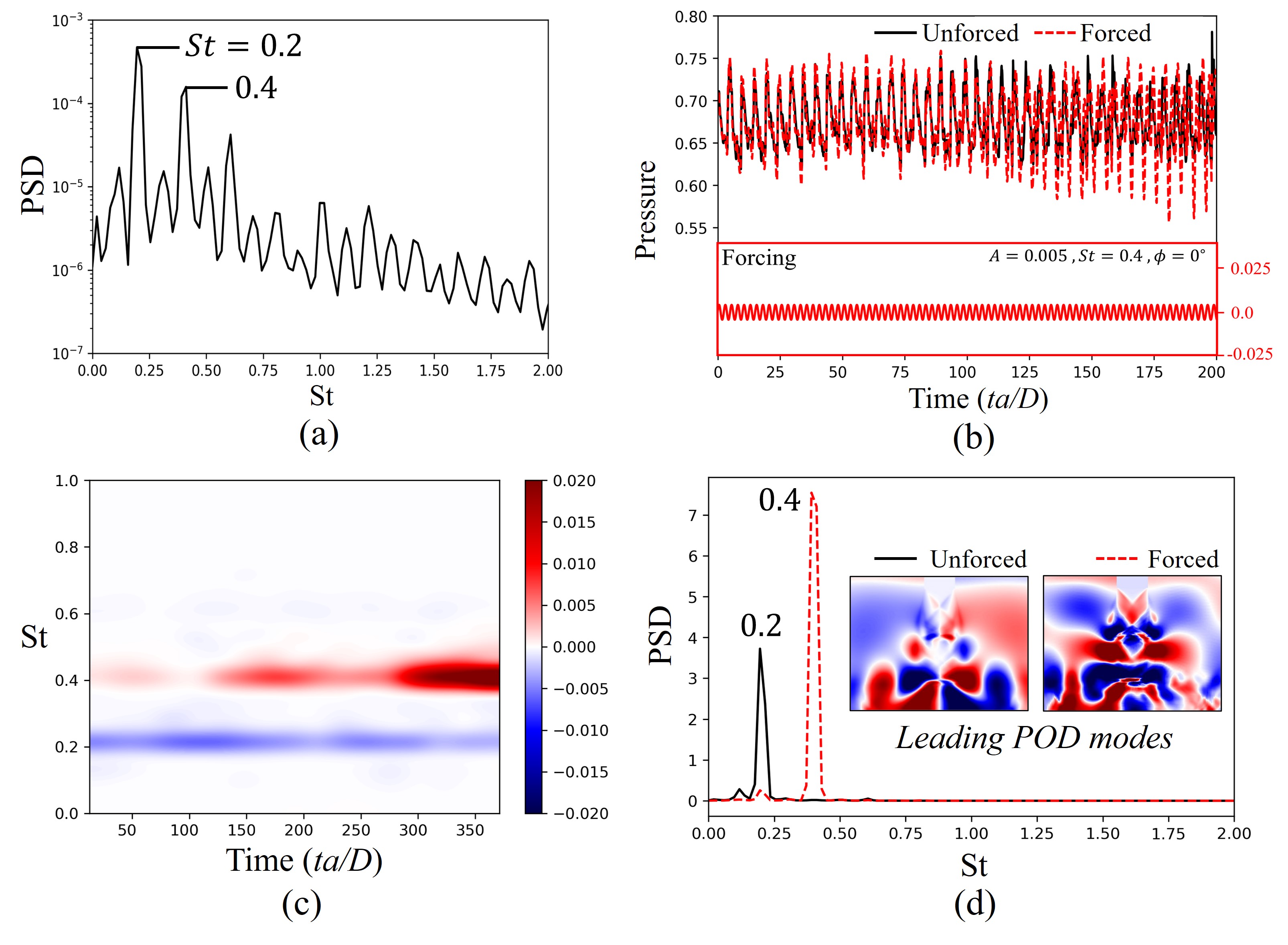}
\vspace*{-0.1in}
\caption{ (a) Unforced acoustic feedback tones at the probe located outside the jet. (b) Initial deviation of the unforced LES and forced DMD-ROM (top) with the forcing input signal (bottom). (c) Scaled DMD-ROM acoustic spectrogram subtracted from unforced spectrogram. (d) Leading POD mode and temporal coefficient spectra of the unforced and forced systems.}
\label{fig:SIJ_probe}
\vspace*{-0.1in}
\end{figure}
The power-spectral density (PSD) of the uncontrolled pressure fluctuations from the probe are plotted in Fig.~\ref{fig:SIJ_probe}(a).
Several tones are evident, the two loudest of which are at $St=0.2$ and $St=0.4$.
These correspond respectively to asymmetric and symmetric patterns, consistent with prior studies \citep{stahl_caes2022}, as elaborated further below.
We consider a broad range of forcing amplitude and frequency parameters.
To illustrate the capability of the DMD-ROM approach, we first discuss in detail the effect of sinusoidal forcing at a frequency of $St=0.4$ and amplitude $A=0.005$ (lower part of Fig.~\ref{fig:SIJ_probe}(b)), which is only  approximately $10\%$ of acoustic fluctuations at the probe.
The forcing is applied to the both sides of the nozzle exit.
These conditions represent the potential manipulation of the resonance mode from the loudest asymmetric tone at $St=0.2$ to the symmetric mode at $St=0.4$. 
A short-time comparison of the unforced and forced DMD-ROM probe signals, top part of Fig.~\ref{fig:SIJ_probe}(c), displays a gradual deviation of the pressure over time due to the forcing. 

The long term, asymptotic influence of the forcing on the resonance tones is displayed in Fig.~\ref{fig:SIJ_probe}(c) with a spectrogram of the probe signals.
To highlight tonal differences, the DMD-ROM results are scaled by the gain ($\sigma=1.44$) and subtracted from the unforced spectrogram; absolute gain behavior is addressed later.
The spectrogram demonstrates that the $St=0.4$ tone increases with time, while the $St=0.2$ tone diminishes.
Changes to the overall flow-field are shown in Fig.~\ref{fig:SIJ_probe}(d) by the leading POD modes of the unforced and forced systems, along with their corresponding unscaled temporal coefficient spectra.
The results demonstrate that the dominant mode without control is asymmetric at $St=0.2$, while that in the forced case is symmetric at $St=0.4$.
Clearly the forcing resonates with the symmetric impinging mode, which overtakes the nominal flapping mode as the leading dynamic. 


To confirm the results from the model, an LES with control is employed by applying perturbations at $St=0.4$ along the spanwise length of the nozzle using a realizable blowing-suction actuator. 
As
depicted in Fig.~\ref{fig:SIJ_intro}(c), actuation is applied on the inner side of the nozzle lip by specifying choked flow based on the main impinging jet nozzle condition.
\begin{figure}
\centering
\includegraphics[width=0.9\textwidth]{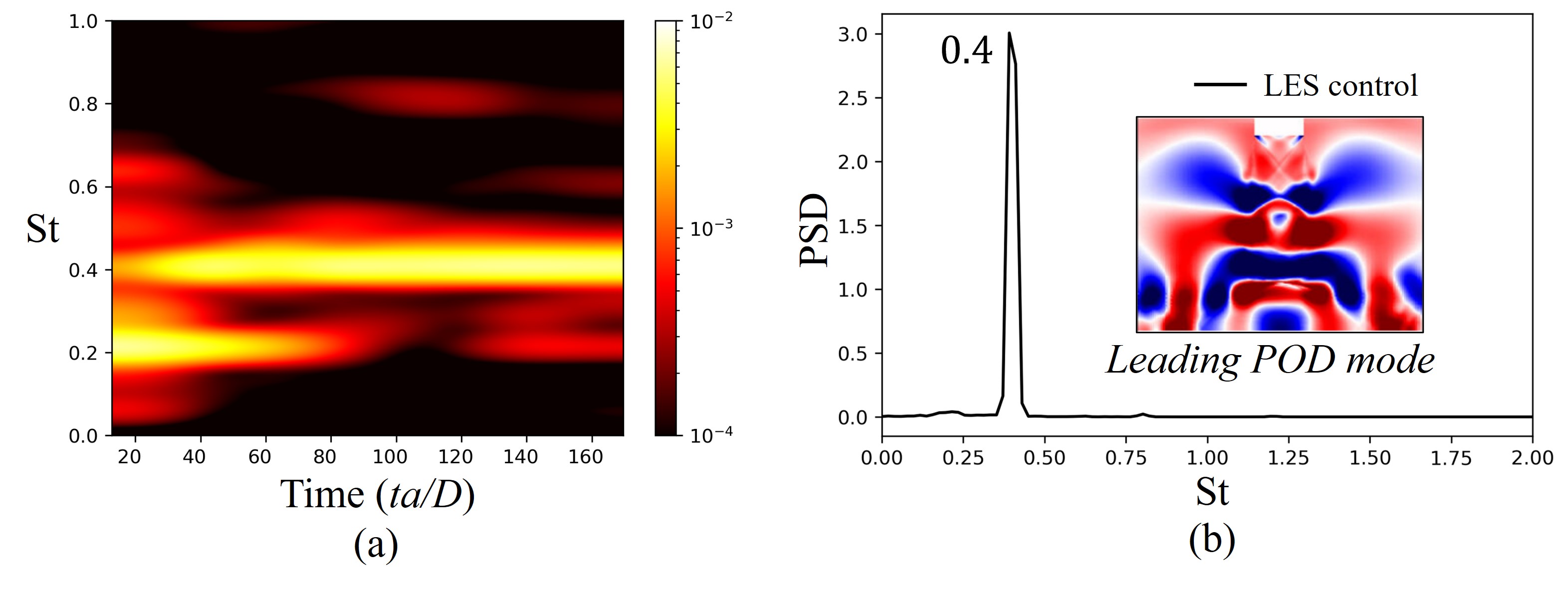}
\vspace*{-0.1in}
\caption{ Validation of the LES blowing-suction at $St=0.4$ showing the (a) controlled LES spectrogram and (b) leading POD mode and temporal coefficient spectra.  }
\label{fig:DMDC_example}
\vspace*{-0.1in}
\end{figure}
The acoustic probe spectrogram from the LES results is presented in Fig.~\ref{fig:DMDC_example}(a).
As in the DMD-ROM, the initially louder $St=0.2$ tone diminishes with time and is replaced by the $St=0.4$ tone as the forcing dominates the solution.
While the frequencies and overall dynamics are predicted favorably, the magnitude of the tones are overpredicted by the DMD-ROM approach.
This is a consequence of its linear nature, which precludes non-linear saturation mechanisms and thus motivated the gain scaling in Fig.~\ref{fig:SIJ_probe}(c) to isolate the observed trends.
The leading POD mode and its PSD obtained from the LES are shown in Fig.~\ref{fig:DMDC_example}(b).
A comparison with POD modes of Fig.~\ref{fig:SIJ_probe}(d), demonstrate a similarity between the controlled LES and DMD-ROM forced results, reflecting the fact that the leading mode is now symmetric, as opposed to the asymmetric unforced case.

\begin{figure}
\centering
\includegraphics[width=0.9\textwidth]{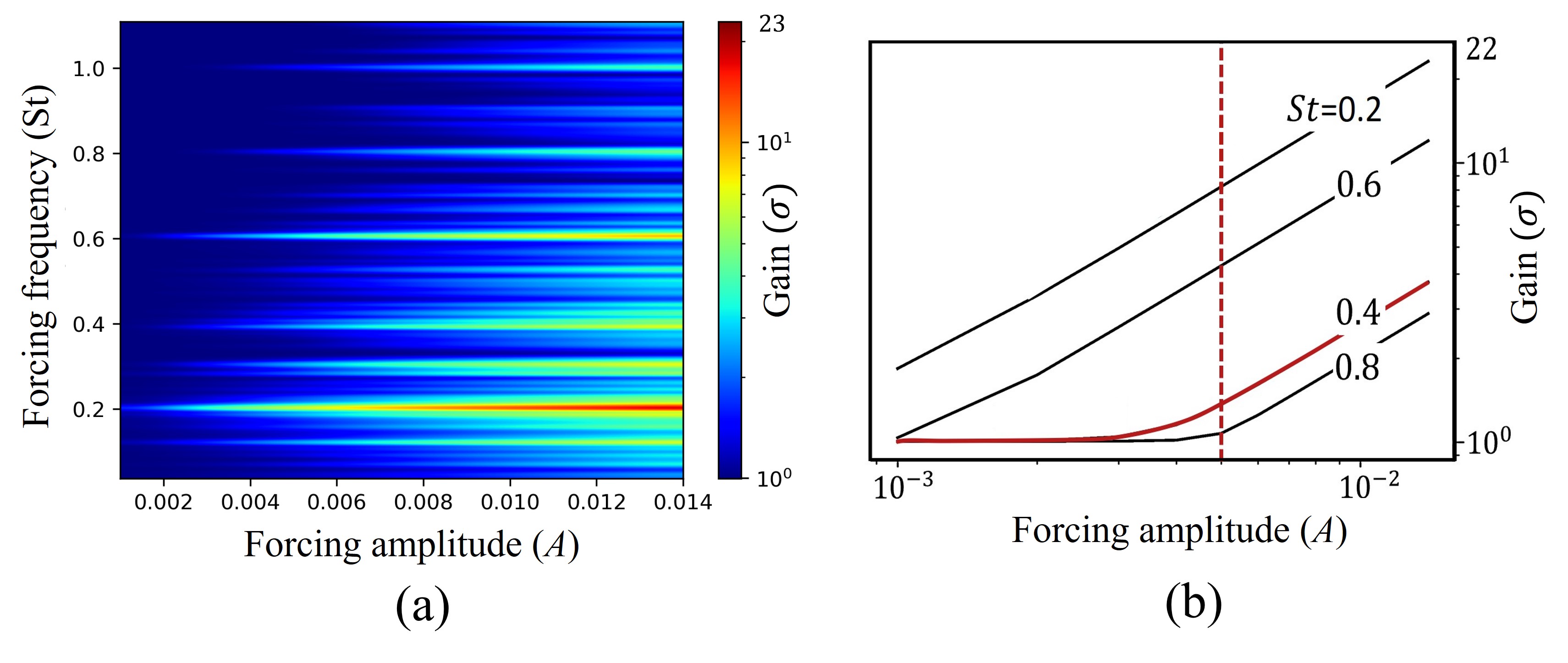}
\vspace*{-0.1in}
\caption{(a) DMD-ROM gain as a function of forcing amplitude and frequency. (b)  Forcing amplitude at  resonance tones.}
\label{fig:subspace}
\vspace*{-0.1in}
\end{figure}

The gain behavior is now examined for a range of forcing frequencies and amplitudes.
Figure~\ref{fig:subspace}(a) shows the gain from $St=0.03$ to $1.1$ and amplitudes $A=0.001$ to $0.015$. 
As noted earlier, for this problem, the gain (\S\ref{secn:method}) measures the ratio of the L2-norm for all forced snapshots to unforced snapshots and therefore captures the asymptotic trends in total energy growth over the time period of the unforced data.
Several streaks are observed, corresponding to frequencies amplified by the flow, among which the resonance tones are particularly susceptible and thus prominent.
Figure~\ref{fig:subspace}(b) selects some of these amplified tones to isolate gain versus forcing amplitude; the previous detailed results at $St=0.4$ forcing are marked with a red line.
For some frequencies, e.g., $St=0.2$ and $0.6$, growth is observed at all amplitudes.  
For others, including $St=0.4$, a larger minimum amplitude is required to observe the initiation of gain, after which the behavior is similar and rises linearly on the log-log scale.
This behavior is attributable to amplitudes that are too low to grow appreciably against the fluctuating snapshots over the finite time period of the ROM. 
Therefore, larger amplitudes are preferred for sensitivity studies to better educe these relative growth rates.
On the other hand, some frequencies in Fig.~\ref{fig:subspace}(a) display no realizable gain ($\sigma=1$) for all amplitudes, indicating these forcing conditions have little influence on the flow.
Of course, forcing characteristics that inhibit pressure fluctuations ($\sigma < 1$) are of great interest from a practical perspective, but the present approach, like other linear methods, highlights only growing modes.
However, the upcoming transient analysis discusses a method to inhibit growth by interfering with time-local events.

\subsection{Example 2: Transient receptivity analysis}\label{secn:examp2}
\begin{figure}
\centering
\includegraphics[width=0.9\textwidth]{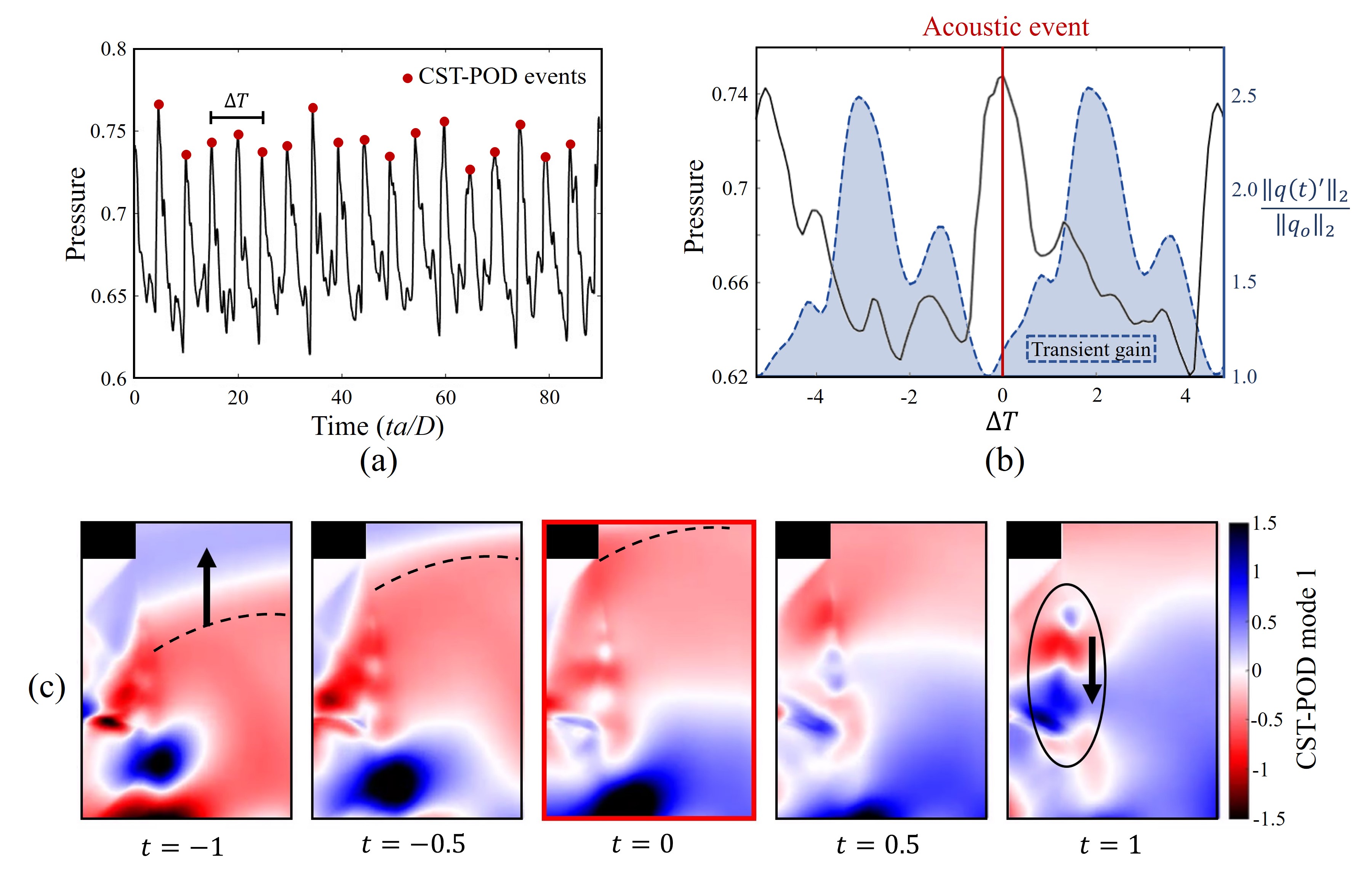}
\caption{(a) Acoustic feedback events from pressure probe. (b) Isolated event from the pressure signal (black) in comparison to the CST-POD transient gain (shaded) through two feedback cycles. (c) CST-POD mode progression in time. }
\label{fig:CPODmode}
\vspace*{-0.1in}
\end{figure}

To demonstrate the versatility of the DMD-ROM approach, the statistically stationary fluctuations of the prior example are replaced by a conditional space-time mode that isolates the sequence of acoustic wave arrival followed by growth of the convective shear-layer instability.
For this, the DMD-ROM is cast as a transient receptivity analysis subject to external forcing, where the gain is measured over a short time-horizon.
The CST-POD calculation is derived as described in \citet{stahl_2023_CPOD_jcp}. 
Briefly, the unforced pressure signal at the probe (shown earlier in Fig.~\ref{fig:SIJ_intro}(a)) is analyzed to conditionally identify finite-time events and corresponding snapshots.
A few peak amplitudes in the pressure signal are shown in Fig.~\ref{fig:CPODmode}(a) for reference, representing acoustic waves passing over the probe.
Each CST-POD event time-window spans $\Delta T=10$ ($100$ snapshots) and is centered between two feedback cycles, capturing $74$ total events across the entire data set to generate the ensemble. 
A sample event is shown in Fig.~\ref{fig:CPODmode}(b) with the probe pressure (left axis) plotted with the black solid curve.

The CST-POD mode is obtained from a local spatial domain near the nozzle where the events are sampled; a few representative snapshots are depicted in Fig.~\ref{fig:CPODmode}(c) to highlight the upstream moving acoustic wave and the ensuing shear-layer instability. 
The transient gain of the CST-POD mode is normalized by the first CST-POD snapshot and is shown by the shaded curve in Fig.~\ref{fig:CPODmode}(b), illustrating the timing between the acoustic event ($t=0$) and peak shear-layer growth ($t=+2$) in the feedback cycle.
This transient gain represents the baseline for evaluating the flow response to different external forcing parameters.



The DMD-ROM forcing for this example case is similar to the previous case, but is only applied on the nozzle side where the CST-POD mode was derived.
Since the CST-POD data 
is not statistically stationary, 
both $\textbf{U}$ and $\tilde{\textbf{A}}$ are calculated from the full variable without subtracting the mean.
In addition to forcing frequency and amplitude, the phase now becomes an important factor because of the short time horizon.
Results using the previous $St=0.4$ forcing with an optimal amplitude and phase are first presented before examining the entire parameter space. 
The choice of the amplitude is arbitrary since it is referenced to the CST-POD mode; in this case $A=0.5$  captures the desired trends in the neighborhood of the local minimum.
The optimal phase is $\phi=150^{\circ}$ based on the parameter sweep presented below.
\begin{figure}
\centering
\includegraphics[width=0.9\textwidth]{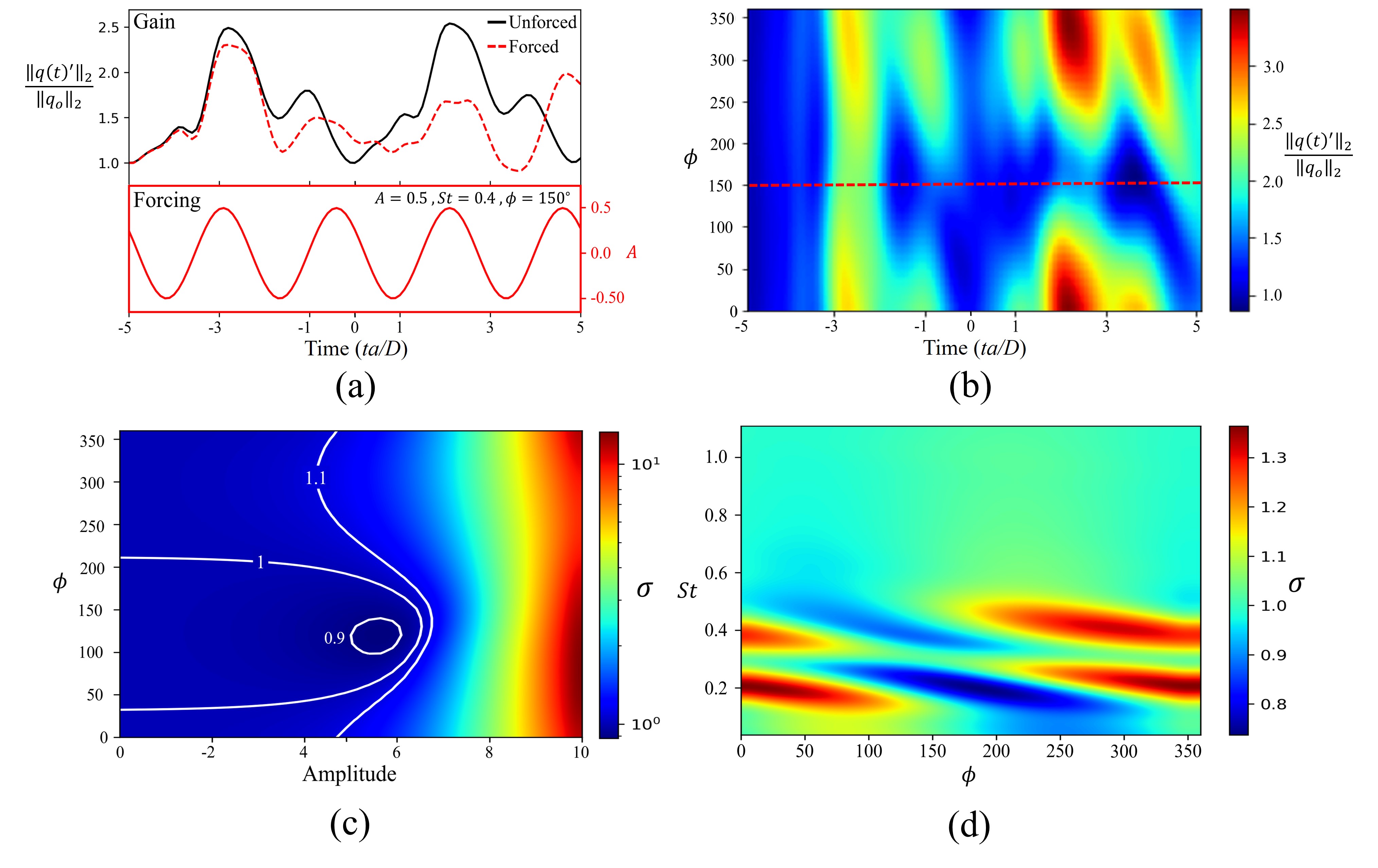}
\vspace*{-0.1in}
\caption{(a) Transient gain of the unforced and forced feedback cycles ($St=0.4$, $A=0.5$, and $\phi=150^{\circ}$). (b) Transient gain as a function of forcing phase. (c) Overall gain as a function of phase and amplitude ($St=0.4$). (d) Overall gain as a function of frequency and phase ($A=0.5$).}
\label{fig:CPOD_time}
\vspace*{-0.1in}
\end{figure}
Figure~\ref{fig:CPOD_time}(a) displays the forcing (lower part of the figure) and gains associated with unforced and forced responses.
The results show that the transient gain is significantly reduced at the peak of the shear-layer instability.
Figures~\ref{fig:CPOD_time}(b-d) illustrate the shear-layer receptivity as a function of forcing phase.
In (b), the transient gain shows phase effects for a constant $St=0.4$ forcing.
At $\phi=150^{\circ}$, the forcing influences the natural instability processes and dampens the peak at $t=+2$.
Figure~\ref{fig:CPOD_time}(c) plots the overall gain ($\sigma$)  as a function of amplitude and phase for $St=0.4$. 
As with the asymptotic case of Example~1, the gain increases linearly with forcing amplitude; this holds for all other frequencies as well. 
However, in this transient case, conditions where gain is reduced ($\sigma<1$) become more apparent and are of interest.
Figure~\ref{fig:CPOD_time}(d) explores the larger parameter space by plotting the overall gain as frequency and phase are varied. 
As expected, the lower frequencies display the largest modulation in feedback instability growth; these are on the order of the resonant frequencies and have fewer cycles within the CST-POD time-window.
In contrast, higher-frequencies have little influence on receptivity and are less affected by phase. 

\section{Discussion}
A data-driven framework  is presented to discern the linear response of a turbulent flow with a view towards control analyses.
The nonlinear system is modeled with a  dynamic mode decomposition reduced order model (DMD-ROM).
Prescribed forcing conditions are projected onto the same reduced space as the unsteady flow, and its effects obtained in the time-domain. 
Computational efficiency is ensured by calculating the gain due to forcing in the reduced subspace, which enables rapid scanning of large ranges of forcing parameters, from which candidates may be expanded to the physical space for further study and verified with scale-resolved methods.

The key advantages of the method are that, being data-driven, it may be applied to any flowfield for which time-resolved  snapshots are available, regardless of geometry and flow parameters, without need to solve the linearized  Navier-Stokes equations. 
Furthermore, the time-domain nature of the technique facilitates examination of the forced evolution in the unsteady flow itself, eases actuator perturbation inputs and examination of transient events where phase becomes important.

Application of the model to the complex physics associated with a resonating supersonic impinging jet forced at the nozzle receptivity region, successfully predicts the switch from flapping to symmetric resonance modes, as validated with a full three-dimensional, nonlinear simulation.
For transient events, the effects of forcing at different phases within the resonant feedback cycle provides insight into means of diminshing or amplifying convective instabilities with respect to the critical events within the cycle.
In both examples, thousands of forcing parameters were tested across amplitude, frequency, and phase  variables, highlighting the performance and utility of the method.

\textbf{Acknowledgments:} The authors acknowledge support from the Collaborative Center for Aeronautical Sciences and the Office of Naval Research.

\textbf{Declaration of interest:} The authors declare no conflict of interest.

\vspace*{-0.1in}
\bibliographystyle{jfm}
\bibliography{Current_DMD_ROM_JFMrapids.bib}

\end{document}